\documentclass[twocolappendix]{emulateapj}

\usepackage[usenames,dvipsnames]{color}
\usepackage{comment}

\shorttitle{Magnetic Fields via Rayleigh-Taylor} 
\shortauthors{Duffell \& Kasen}

\begin{document}

\title{Synchrotron Magnetic Fields from Rayleigh-Taylor Instability in Supernovae}

\author{Paul C. Duffell and Daniel Kasen}
\affil{Astronomy Department and Theoretical Astrophysics Center, University of California, Berkeley, CA 94720}
\email{duffell@berkeley.edu}

\begin{abstract}

Synchrotron emission from a supernova necessitates a magnetic field, but it is unknown how strong the relevant magnetic fields are, and what mechanism generates them.  In this study, we perform high-resolution numerical gas dynamics calculations to determine the growth of turbulence due to Rayleigh-Taylor instability, and the resulting kinetic energy in turbulent fluctuations, to infer the strength of magnetic fields amplified by this turbulence.  We find that Rayleigh-Taylor instability can produce turbulent fluctuations strong enough to amplify magnetic fields to a few percent of equipartition with the thermal energy.  This turbulence stays concentrated near the reverse shock, but averaging this magnetic energy throughout the shocked region (weighting by emissivity) sets the magnetic fields at a minimum of 0.3 percent of equipartition.  This suggests a minimum effective magnetic field strength ($\epsilon_B > 0.003$) which should be present in all interacting supernovae. 

\end{abstract}


\section{Introduction} \label{sec:intro}

The interaction of the ejecta from a supernova explosion with a surrounding circumstellar medium (CSM) can give rise to synchrotron radiation at radio wavelengths.  Radio observations of core collapse supernovae (SNe) have been used to constrain the density and structure of the CSM, with implications for the presupernova evolution and mass loss history of massive stars \citep[e.g.,][]{2006ApJ...651..381C, 2007Sci...317..924P}.  Radio non-detections in Type Ia SNe have been used to place some of the strongest constraints on the progenitor system.  Upper limits from deep radio images set bounds on the CSM density from the stellar winds that are expected from certain binary channels \citep{2012ApJ...752...78S, 2012ApJ...746...21H, 2012ApJ...750..164C, Chomiuk2015}.  

Perhaps the largest current uncertainty in physically interpreting the radio emission from
interacting SNe is the strength of the magnetic field in interaction region.  Essentially all synchrotron modeling to date assumes that the field strength can be characterized by a constant parameter $\epsilon_B$, the ratio of magnetic to thermal energy in the flow.  The value of $\epsilon_B$ is not well-constrained by theory, as it is unclear what mechanism is responsible for the field's amplification in supernova shocks.  Observation can only set broad limits, as the value of $\epsilon_B$ is degenerate with other properties of the flow.  The large uncertainty on $\epsilon_B$ makes it difficult to make strong inferences from radio non-detections.  If there is no minimum ``floor" on $\epsilon_B$, then some outflows could have arbitrary small magnetic field, and hence be undetectable in the radio, even if the CSM density is high.

Theoretical constraints on $\epsilon_B$ are difficult to make, in part because the underlying cause of the magnetic fields is uncertain.  If the magnetic field strength, $B$, in the supernova is inherited from the progenitor, it will drop quickly as the flow expands via flux freezing: $B \sim B_0 (R(t)/R_0)^{-2}$, where $B_0$ is the surface magnetic field of the progenitor, $R(t)$ is the shock radius, and $R_0$ is the progenitor size.  Assuming a solar-like progenitor with $\sim 1$ Gauss magnetic field, CSM density of $\rho \sim A/r^2$ with $A = 5 \times 10^{11}$ g/cm and ejecta velocity of $10^9$ cm/s, this gives a meager $\epsilon_B \sim B^2 / (\rho_{\rm CSM} v^2) \sim 10^{-8} (R/R_0)^{-2}$, where at typical times of observation, $t \sim 1$~day, the  radius $R = vt \gg  R_0$.  On the other hand, plasma instabilities such as Weibel instability \citep{1999ApJ...511..852G, 1999ApJ...526..697M} have been shown to amplify magnetic fields to $\epsilon_B \sim 10\%$  in the context of relativistic jets \citep{2003ApJ...595..555N, 2008ApJ...682L...5S}.  However, it is expected that this field should only be present within a few plasma skin depths of the forward shock, resulting in a negligible emitting volume for non-relativistic SN interactions.

While the prospects of calculating an $\epsilon_B$ from theory have previously seemed remote, we argue here that a robust minimal value can be estimated from straightforward hydrodynamical calculations.  Interacting SNe are turbulent due to Rayleigh-Taylor (RT) instability, so a magnetic field in the flow is necessarily set by small scale turbulent dynamo.  We will show numerically that a minimal floor to $\epsilon_B$ is set by this process, even if all other processes (e.g. Weibel instability) fail to generate significant magnetic fields. 

Amplification of magnetic fields by RT was first studied by \cite{1995ApJ...453..332J} in two and three-dimensions with magnetic fields using local calculations.  \cite{1996ApJ...465..800J} demonstrated in the supernova context (in 2D) that RT could cause these fields to align with turbulent structures, affecting polarization of synchrotron emission.  \cite{2000ApJ...528..989K} studied the difference between 2D and 3D, but still in a local sense (looking at single-mode perturbations).  That study found that the growth of RT is $30-35\%$ stronger in 3D than in 2D.  Magnetic amplification due to RT and small-scale turbulent dynamo has also been studied in the relativistic case, in the context of gamma ray bursts \citep{2013ApJ...775...87D, 2014ApJ...791L...1D}.

Studying the entire process self-consistently in a global calculation is computationally demanding.  First, a proper treatment of MHD turbulence necessitates a 3D calculation.  Secondly, capturing small-scale turbulent dynamo in MHD turbulence requires high resolution; at least 256 zones across the largest-scale eddies \citep{2013ApJ...769L..29Z}.  This is a known issue, for example, in simulations of merging neutron stars, in which small-scale turbulent dynamo is expected to amplify magnetic fields up to magnetar-levels \citep{2013ApJ...769L..29Z, 2015ApJ...809...39G}.  Global simulations, however, have yet to observe this dramatic amplification (but they have come close; see \cite{2015arXiv150909205K}).

Rather than attempt a full 3D MHD study which resolves the dynamo, we restrict ourselves to modest 2D hydrodynamics calculations, to determine the amplitude of turbulent fluctuations in the saturated state.  We then use this to infer the strength of the magnetic fields as predicted by dynamo theory, and the consequences for the radio emission from SNe.  Although a full 3D calculation would be more accurate in many ways (for example, the 2D solution will not have the correct kinetic power spectrum), the 2D calculation will capture the strength of the driving field to order-of-magnitude \citep[as shown by previous studies comparing 2D and 3D RT, e.g.][]{2000ApJ...528..989K}, and this large-scale driving field is what sets the turbulent kinetic energy.

\section{Numerical Set-Up} \label{sec:numerics}

\begin{figure*}
\epsscale{1.15}
\plotone{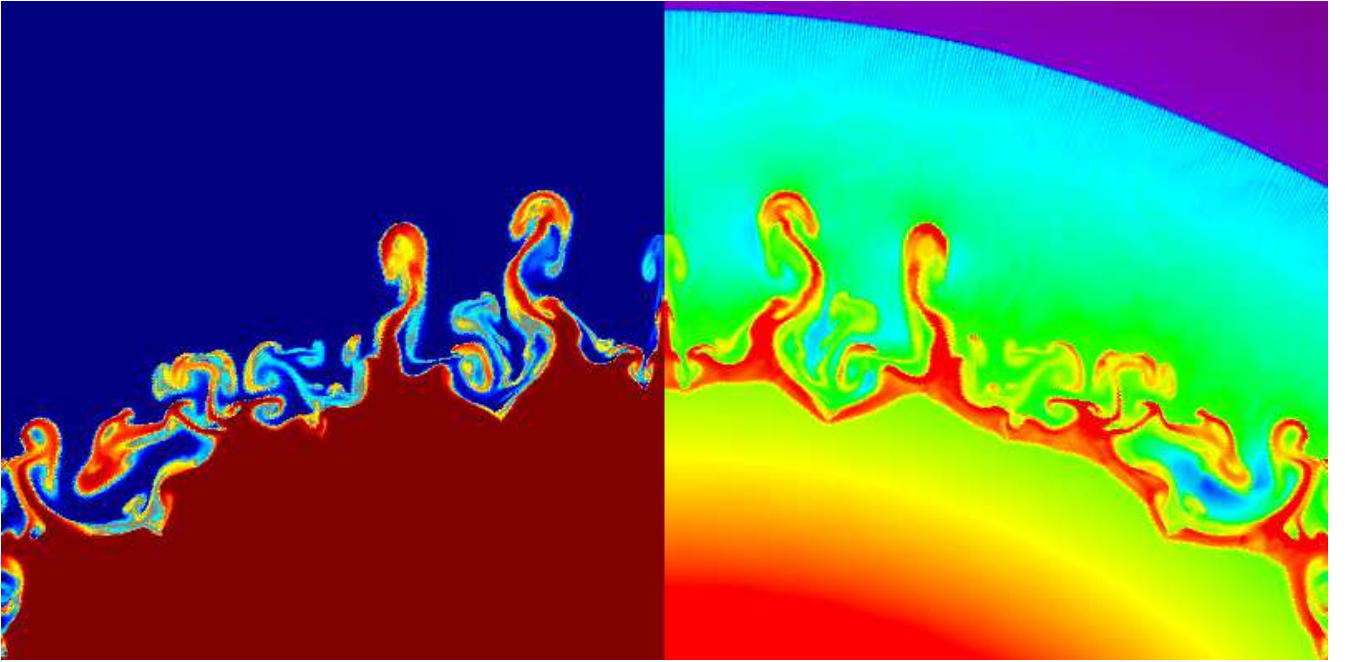}
\caption{ Rayleigh-Taylor instability in the case with very small seed density fluctuations $\delta_0 = 0.01$ (compare Figure 6 and 7 of \cite{1992ApJ...392..118C}).  The left side plots the passive scalar $X$ (ranging between $0$ and $1$, with $0$ for pure ejecta and $1$ for pure CSM).  The right side plots the logarithm of density (with dimensions scaled out, and ln($\rho$) ranging from $1$ to $4$).
\label{fig:pretty} }
\end{figure*}

\begin{figure}
\epsscale{1.15}
\plotone{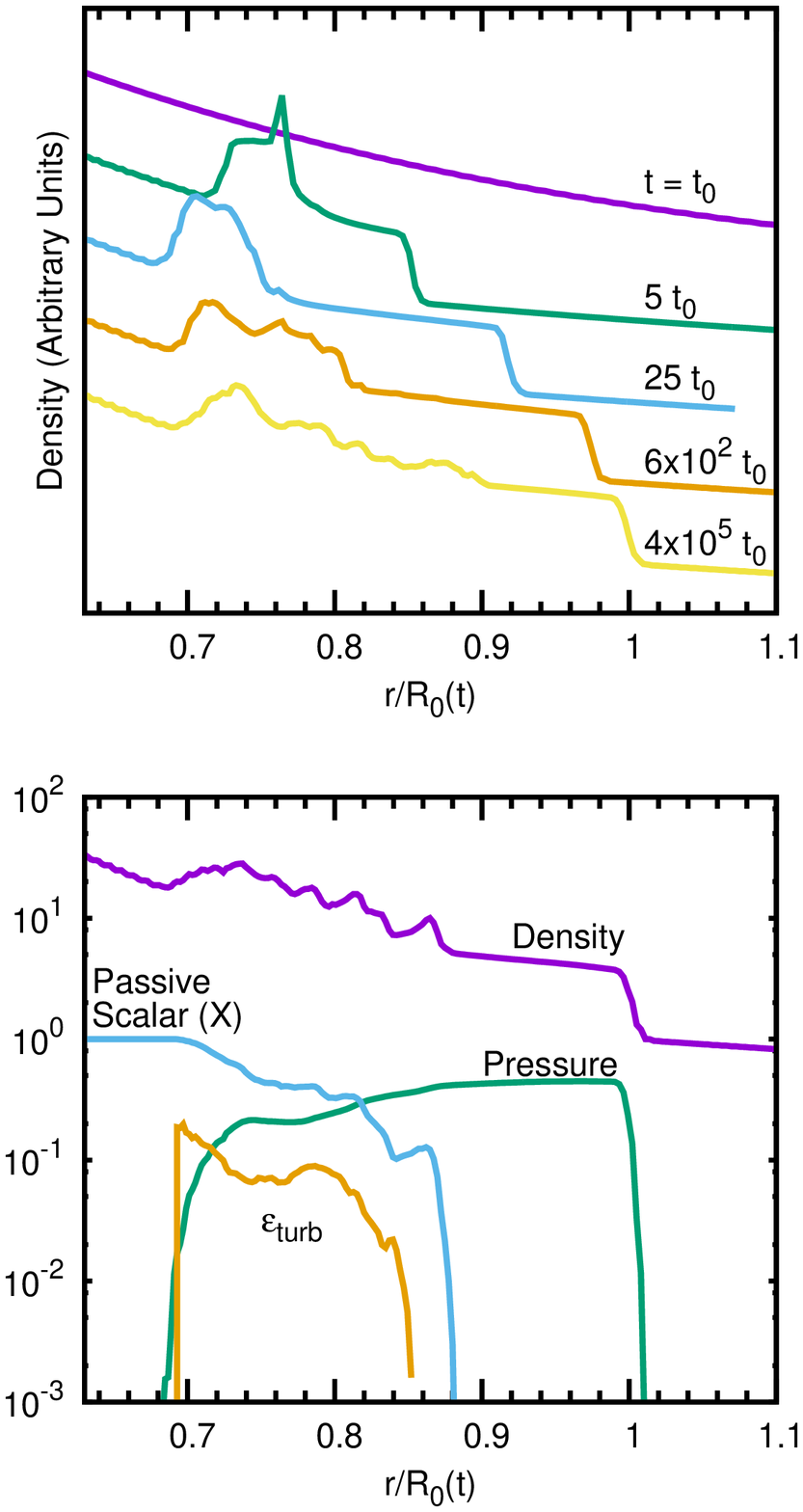}
\caption{ \emph{Upper Panel:}  Spherically-averaged profiles of density as a function of (re-scaled) radius at various times during the evolution.  Density is plotted on an arbitrary scale, multiplied by an offset to facilitate comparison.  After a time around $t \sim 100 t_0$, the flow has attained a (statistically) self-similar state.  \emph{Lower Panel:}  Profiles of density, pressure, $X$, and $\epsilon_{\rm turb}$ in this self-similar solution.  $\epsilon_{\rm turb}$ attains values of order $0.1$, but is concentrated near the reverse shock.
\label{fig:1d} }
\end{figure}

The numerical calculations performed in this study are very similar to those performed by \cite{1992ApJ...392..118C},  who studied RT instability in interacting SNe.  We focus on measuring the kinetic energy density of turbulent fluctuations and mapping this to a magnetic field strength.

Our numerical calculations  integrate the equations of two-dimensional (2D) axisymmetric hydrodynamics

\begin{equation} 
\partial_t ( \rho ) + \nabla \cdot ( \rho \vec v ) = 0,
\end{equation} 
\begin{equation}
\partial_t ( \rho v_r ) + \nabla \cdot ( \rho v_r \vec v + P \hat r ) = ( 2 P + \rho v_\theta^2 )/ r,
\end{equation} 
\begin{equation}
\partial_t ( r \rho v_\theta ) + \nabla \cdot ( r \rho v_\theta \vec v + P \hat \theta ) = P cot \theta,
\end{equation} 
\begin{equation}
\partial_t ( \frac12 \rho v^2 + \epsilon ) + \nabla \cdot ( ( \frac12 \rho v^2 + \epsilon + P ) \vec v ) = 0,
\end{equation} 
where $\rho$ is density, $P$ is pressure, $\epsilon$ is the internal energy density, and $\vec v$ is the velocity.  The equation of state is assumed to be gas pressure dominated: $\epsilon = \frac{3}{2} P$.  This is appropriate for interaction with a sufficiently low-density CSM, for which the timescale for the shocked gas to radiate is much longer than the dynamical timesale.  A few additional runs were performed with variable adiabatic index, to determine the dependence of the results on the equation of state.

Numerical calculations are carried out using the JET code \citep{2011ApJS..197...15D, 2013ApJ...775...87D}, a moving mesh technique that is effectively Lagrangian due to the radial motion of computational zones.
The initial conditions at the start time $t_0$ adopt a power-law profile for both the ejecta density, $\rho_{\rm ej}$, and the CSM density, $\rho_{\rm CSM}$,
\begin{equation}
\rho_{\rm ej}(r,t_0) = \left( { r \over g t_0 } \right)^{-n} t_0^{-3},
\end{equation} 

\begin{equation}
\rho_{\rm CSM}(r,t_0) = q r^{-s},
\end{equation} 
where $g$ and $q$ are constants that set the density scales.
For the present study, we choose $n=7$, $s=2$.  In a similar study, \cite{1992ApJ...392..118C}  use power-law profiles for the ejecta and CSM structures, but initialized the contact region with the self-similar solutions of \cite{1982ApJ...258..790C}.  In contrast, this study allows the ejecta and CSM to collide and evolve to the self-similar structure.  We find that RT instability sets in before the self-similar Chevalier solution has had time to emerge.

Rather than initializing the contact interface between the ejecta and CSM as a step function, it is numerically advantageous to start with a smooth initial condition

\begin{equation}
\rho_{\rm tot}(r,t_0) = \rho_{\rm ej}(r,t_0) + \rho_{\rm CSM}(r,t_0).
\end{equation} 

This initial condition is asymptotically identical to Chevalier's (for both $r \rightarrow 0$ and $r \rightarrow \infty$), and so at late times in 1D it is guaranteed to approach the same self-similar solution.
 
Additionally, a seed perturbation is introduced, assuming some ``clumpiness" to the ejecta and CSM

\begin{equation}
\rho(r,t_0) = \rho_{\rm tot}(r,t_0)e^{\delta(\vec r)},
\end{equation} 
where the fluctuations $\delta(\vec r)$ are given by

\begin{equation}
\delta(\vec r) = \delta_0 \text{sin}( l ~\theta ) \text{sin}( l ~\text{ln}(r) ),
\end{equation} 
with an angular wavenumber $l$ (where $l = 50$ in this study), and a magnitude  $\delta_0$.  The ejecta is assumed to be cold, ballistic, and expanding homologously, with a velocity profile

\begin{equation} 
v( r , t_0 ) = \left\{ \begin{array}
				{l@{\quad \quad}l}
				r/t_0 & r < R_0(t_0)	\\  
    			0 & r > R_0(t_0) 		\\ 
    			\end{array} \right.    
\end{equation}
where $R_0(t) = (q g^n)^{1 \over n-s}t^{n-3 \over n-s}$ is the initial contact interface.
However, again it is numerically convenient to adopt smooth initial conditions, and therefore we initialize the velocity as
\begin{equation}
v(r,t_0) = (r/t_0) { \rho_{\rm ej}(r,t_0) \over \rho_{\rm tot}(r,t_0) }.
\end{equation} 

Again, as this velocity profile is asymptotically identical to Chevalier's at large and small radii, it approaches an identical solution at late times, after the forward and reverse shocks have swept up a sufficient amount of mass.

Additionally, a passive scalar is initialized in the flow, to differentiate ejecta from CSM, and in principle measure the mixing of the two:

\begin{equation} 
X( r , \theta , 0 ) = \left\{ \begin{array}
				{l@{\quad \quad}l}
				0 & r < R_0(t_0) 	\\  
    			1 & r > R_0(t_0) 	\\ 
    			\end{array} \right.    
\end{equation}

Note that this passive scalar is initialized as a discontinuous function (as opposed to the initial density and velocity) so that values of X that deviate from $1$ or $0$ are entirely due to mixing.  This makes it possible to precisely track which regions of the solution are mixed.

These initial data are then evolved for seven orders of magnitude in time, until $t = 10^7 t_0$.  This ensures that the RT instability has reached a saturated, statistically self-similar state (Figure \ref{fig:pretty}).

\subsection{Turbulence Measurements and Magnetic Field Amplification}

RT instabilities in the interaction region drive turbulent fluctuations that will amplify magnetic fields.  We quantify the magnitude of turbulent fluctuations by measuring the amount of kinetic energy in a spherical shell, then subtracting off the part  attributed to bulk motion.  The total hydrodynamical energy density is a combination of bulk kinetic energy, turbulent kinetic energy, and thermal energy:

\begin{equation}
U_{\rm tot} = \frac12 \rho \left<v\right>^2 + U_{\rm turb} + P/(\gamma-1),
\end{equation}
where $\gamma = 5/3$ is the adiabatic index.  The bulk velocity $\left<v\right>$ is calculated by dividing the spherical shell's momentum by its mass.  Pressure and density are evaluated using a simple volume average over the shell.  Since all other quantities (including total energy) are known for a given spherical shell, $U_{\rm turb}$ can be found by solving this equation.  Then the turbulent energy fraction $\epsilon_{\rm turb} = U_{\rm turb} / U_{\rm thermal}$ can be readily calculated

\begin{equation}
\epsilon_{\rm turb} = {U_{\rm tot} - \frac12 \rho \left<v\right>^2 - P/(\gamma-1) \over P/(\gamma-1) }.
\end{equation}

The quantity $\epsilon_{\rm turb}$ is a measurement of turbulent kinetic energy.  Here we argue that $\epsilon_{\rm turb} \sim \epsilon_{B}$, i.e. that magnetic fields can be quickly amplified by small scale turbulent dynamo up to rough equipartition with the turbulent fluctuations, regardless of the initial seed field.

Equipartition is established by turbulent dynamo processes, which are known to be present in low-viscosity conducting fluids with high magnetic Prandtl number and a persistent injection of turbulent kinetic energy.  Initially, kinematic small-scale turbulent dynamo drives exponential growth of any weak pre-existing magnetic field at a rate comparable to the turnover time of smallest eddies \citep{kazantsev1968enhancement, 1978mfge.book.....M}.  This rate is extremely fast compared to outer-scale eddy turn-over times (in the high Reynolds number limit, this process is effectively instantaneous, which is why it is independent of the initial magnetic field strength).

The kinematic process terminates when the energy in viscous scale magnetic fluctuations balances kinetic energy of viscous scale eddies. Magnetic field amplification then continues via nonlinear small-scale dynamo process \citep{2004ApJ...612..276S}.  During the nonlinear phase, the magnetic energy grows linearly with time, as magnetic fluctuations move to progressively larger scales.  The nonlinear phase terminates when magnetic fluctuations exist in scale-by-scale equipartition up to the outer scale of the RT-inspired turbulence. 

Numerical calculations in the non-relativistic \citep{2003ApJ...597L.141H, 2012PhRvL.108c5002B} and relativistic \citep{2013ApJ...769L..29Z} cases indicate that in the limit of large Reynolds number, non-linear small-scale turbulent dynamo saturates universally after several large-scale turnover times.  Therefore, if magnetic fields and 3D resolution of sufficient turbulent sub-scales were included in our calculations, these magnetic fields should quickly end up in kinetic equipartition with turbulent fluctuations.  ``Equipartition" here means that the ratio of kinetic to magnetic energy $\epsilon_{\rm turb} / \epsilon_B$ is of order unity.  The aforementioned numerical studies have reported end-state turbulent dynamo saturation with magnetic energy at between 30\% and 60\% of the turbulent kinetic energy density.  Therefore, assuming the largest eddies turn over several times during the evolution, $\epsilon_B \approx 0.3 ~\epsilon_{\rm turb}$ at the very least.

\section{Results} \label{sec:results}

Figure \ref{fig:pretty} shows fully-developed turbulence after the flow has expanded by five orders of magnitude (here the seed density fluctuations are $\delta_0 = 0.01$).  The turbulence has grown to large scales, and (as Figure \ref{fig:1d} shows) the coarse-grained properties of the flow are asymptoting toward a (statistically) self-similar solution.  Note that this solution is distinct from the 1D self-similar solution of Chevalier, yet it still appears to obey the same scaling and self-similarity in the angle-averaged profile (scale invariance is not violated by allowing RT-induced mixing).

The top panel of Figure \ref{fig:1d} shows the 1D averaged density, as it approaches statistical self-similarity, and the lower panel shows the late-time solution.  Also shown in the lower panel is the quantity $\epsilon_{\rm turb}$, which takes on values as large as $10\%$ and is concentrated in the mixed region just downstream of the
reverse shock.   Because turbulent mixing does not propagate all the way out to the forward shock,  $\epsilon_{\rm turb}$  falls to zero at larger radii.   
The  magnetic fields generated as a result of RT instabilities will therefore not uniformly fill the shocked region.

Most  analyses of radio supernova observations have assumed a value of $\epsilon_B$ that is constant with radius.    To compare to these studies, we 
define an ``effective  $\epsilon_B$'' for our solutions that involves a weighted average of $\epsilon_{\rm turb}$ over the shocked region.  We begin with the synchrotron power radiated per unit volume between frequencies $\nu$ and $\nu + d\nu$ 
\begin{equation}
p_\nu d\nu =  \frac{4}{3}  \sigma_T c u_B \gamma^2 n(\gamma) d\gamma,
\label{eq:synch_p}
\end{equation}
where $\sigma_T$ is the Thompson cross-section, $c$ is the speed of light, $u_B = B^2/8 \pi$ is the magnetic energy density, and $n(\gamma)$ is the number density of non-thermal electrons as a function of their Lorentz factor $\gamma$.  We assume that  shock acceleration generates a standard power-law distribution, $n(\gamma) = C \gamma^{-p}$ for
$\gamma > \gamma_{\rm min}$.  The constant
 $C = (p-1) n_{\rm nt} \gamma_{\rm min}^{p-1}$ where  $n_{\rm nt}$ is the total number density of non-thermal electrons.  In our study we take $p = 2.5$.
  Assuming that each electron radiates at its characteristic synchrotron frequency, $\nu_c = \gamma^2 e B/m_e c$, we can substitute into Eq.~\ref{eq:synch_p} and integrate over
the shocked volume to get the specific luminosity
\begin{equation}
L_\nu =  \int \frac{2}{3} C  \sigma_T c   u_B \nu_c^{-1} \left( \frac{\nu}{\nu_c} \right)^{(1-p)/2} 
4 \pi r^2 dr.
\end{equation}
We assume that $n_{\rm nt}$ is a fixed fraction of the total gas density, $n_{\rm nt} \propto \rho$, and that the  energy density in non-thermal electrons is a fixed fraction of the total gas energy density, $u_{\rm nt} \propto P$.
%After substituting and pulling out constants \citep[see for example][]{2010ApJ...722..235V}, and also 
Making the assumption $\epsilon_{B} \propto \epsilon_{\rm turb}$, the integral can be expressed as 
\begin{equation}
L_\nu = {\rm const} \cdot \int \epsilon_{\rm turb}^{(p+1)/4} F(r) dr,
\end{equation}
where constants have pulled out of the integral, and $F(r)$ is defined as

\begin{equation}
F(r) = r^2 \rho (P/\rho)^{p-1} P^{(p+1)/4}.
\end{equation}
The equivalent uniform $\epsilon_{\rm turb}$ can now be calculated by an averaging process, weighting by $F(r)$

\begin{equation}
\left< \epsilon_{\rm turb}(t) \right> = \left( \int (\epsilon_{\rm turb}(r))^{p+1 \over 4} F(r) dr \over \int F(r) dr \right)^{4 \over p+1}
\end{equation}
The quantity $\left< \epsilon_{\rm turb}(t) \right>$ is an ``effective average", in the sense that it is equivalent to the $\epsilon_B$ that would be inferred if it were assumed that $\epsilon_B$ was uniform over the entire shocked region.

\begin{figure}
\epsscale{1.15}
\plotone{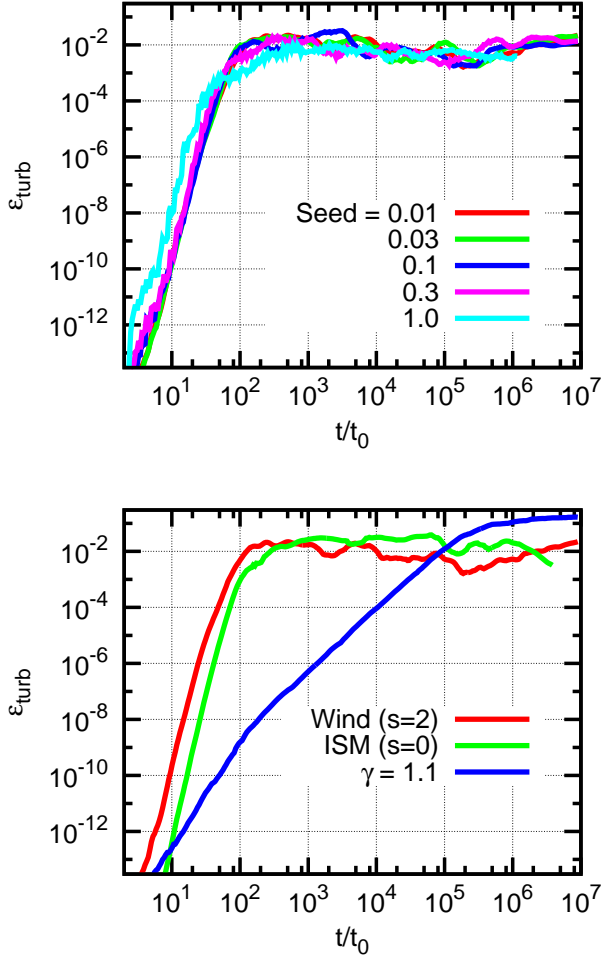}
\caption{ \emph{Upper Panel:} Turbulent energy fraction $\left< \epsilon_{\rm turb} \right>$ as a function of time for different magnitudes of seed perturbation.  The turbulence is nearly independent of seed field, except for the case of order-unity perturbations $\delta_0 = 1.0$.  For all cases, turbulent fluctuations grow as $\sim t^8$, and saturation is established around $t \sim 100~t_0$.  The turbulent energy fraction finds a final saturated state at $\epsilon_{\rm turb} \sim 0.01$. \emph{Lower Panel:} Solution is tested for different hydrodynamical models.  Choosing a uniform CSM (``ISM Model") results in a very similar progression for the turbulent amplification.  Reducing the adiabatic index results in a much larger saturation value of $\epsilon_{\rm turb} \sim 0.3$, but saturation occurs much later, around $t \sim 10^5~t_0$.  This suggests that cosmic ray cooling could affect these results, potentially enhancing $\epsilon_B$, but also affecting the timescale for saturation.
\label{fig:epsB} }
\end{figure}

Figure \ref{fig:epsB} plots this averaged $\left< \epsilon_{\rm turb} \right>$ as a function of time for various choices of $\delta_0$.  Regardless of the seed field, a common picture emerges:  $\left< \epsilon_{\rm turb} \right>$ grows quickly, roughly as $t^8$, until saturating around $\left< \epsilon_{\rm turb} \right> \sim 0.01$ after time $t \sim 100 ~t_0$.  This growth appears to be independent of the magnitude of seed perturbations, so long as such perturbations are below order-unity.

The growth and saturated value of $\left< \epsilon_{\rm turb} \right>$ is largely independent of the seed field, and after time-averaging the saturation level is found to be $\left< \epsilon_{\rm turb} \right> \approx .0097$.  Assuming magnetic fields are amplified to roughly $30 - 60\%$ of this, we find $\epsilon_B \sim .003$ at a minimum due to RT alone.

Results are also included in Figure \ref{fig:epsB} for an additional ``ISM" model, where the CSM density is uniform ($s=0$, lower panel).  This resulted in very similar behavior, suggesting that the strength of the turbulence is not sensitive to the detailed initial setup.

As a first attempt at modeling the effects of cooling, we perform an additional calculation with a soft equation of state, using an adiabatic index $\gamma = 1.1$, rather than the $\gamma = 5/3$ value used primarily in this study.  In the lower panel of Figure \ref{fig:epsB}, $\epsilon_{\rm turb}(t)$ is shown for the $\gamma = 1.1$ calculation, showing that softening the equation of state  impacts the results.  Because the shocked region is narrower for lower $\gamma$, turbulence catches up to the forward shock (as noticed by \cite{2001ApJ...560..244B}).  Nearly the entire shocked region becomes turbulent and magnetized, increasing the emitting volume and therefore the ``effective average" of $\epsilon_{\rm turb}$.  We find saturation at $\epsilon_{\rm turb} \sim 0.3$, suggesting a floor of $\epsilon_B \sim 0.1$.  However, the timescale for reaching saturation is also affected by the equation of state, causing saturation as late as $10^5 t_0$.  A more detailed calculation including realistic models for cooling is warranted, and will be attempted in a future study.

\section{Discussion} \label{sec:disc}

RT instability generates turbulence in interacting supernovae, which is amplified until reaching a saturated state after time $t \sim 100~t_0$, where $t_0$ is the time when the encounter with the ambient medium begins: $t_0 \sim R_0/v_0$, where $R_0$ is the inner radius of the CSM and $v_0$ is the ejecta velocity.  This timescale for saturation is in agreement with that found by \cite{1992ApJ...392..118C}.  For example, assuming observed ejecta velocities of $v_0 \sim 10^9$ cm/sec, and a CSM around a solar-size progenitor, $R_0 \sim 10^{11}$ cm, the saturated turbulent state is reached after $100 t_0 \sim$ hours.  For CSM around a larger progenitor (such as a red supergiant), $100 t_0 \sim$ days, while for a more compact progenitor (as  a white dwarf in a type-1a supernovae) $100 t_0 \sim$ minutes.  These numbers are independent of the clumpiness of the surrounding medium, so long as such clumpiness consists of less than order-unity perturbations to the density. 

If RT generated turbulence is the only mechanism producing magnetic fields in interacting supernovae, then one would expect to see significantly less emission prior to $t \sim 100 t_0$.  Depending on the distribution of CSM, this may have a significant impact on the predicted light curves of radio supernovae.  The initial stages of the interaction may be rendered radio invisible, as the magnetic fields have not had time to be amplified.

Assuming magnetic energy is amplified to equipartition with kinetic fluctuations, the saturated magnetic field in the asymptotic state is $\epsilon_B \sim 3\%$ of equipartition with the thermal energy in the vicinity of the reverse shock.  If no other mechanism produces sufficient magnetic fields, the reverse shock  dominates the synchrotron emission, which is in contrast to standard models assuming a constant $\epsilon_B$, where the reverse shock contributes only $\sim 10\%$ of the total luminosity.  A value of $\epsilon_B \sim 3\%$ near the reverse shock translates to an effective averaged magnetic energy of $\epsilon_B \sim 0.3\%$; that is, it generates as much synchrotron flux as a supernova with uniform $\epsilon_B \sim 3 \times 10^{-3}$ everywhere.  Therefore, applying the value $\epsilon_B \sim 3 \times 10^{-3}$ is justified in that it is a lower bound to the magnetic field strength which should be present in all supernovae from Rayleigh-Taylor instability alone.

These numbers may be enhanced by cosmic ray cooling, as it has a significant impact on the dynamics of RT \citep{2001ApJ...560..244B}.  Cosmic rays provide significant cooling in shocks, slowing down the forward shock, and reducing the steepness of pressure gradients, allowing the Rayleigh-Taylor fingers to propagate through the entire shocked region.  This was demonstrated first by \cite{2001ApJ...560..244B}, by varying the adiabatic index, as an effective proxy for cooling.  This has also been shown in the relativistic case \citep{2014ApJ...791L...1D}.  \cite{2010A&A...515A.104F} and \cite{2010A&A...509L..10F} have also seen this effect, using a prescribed model for cosmic ray cooling (rather than varying the adiabatic index).  The importance of cosmic rays on the dynamics has also been shown observationally \citep{2005AAS...20717212W}.  Our studies using a softer equation of state demonstrate this trend, suggesting that the floor on $\epsilon_B$ is even larger than the conservative $0.3\%$ minimum calculated here (though it appears the softer equation of state affects the timescale for saturation, see Figure \ref{fig:epsB}).  A full calculation including these effects (and the effects of 3D turbulence) will be attempted in a future study.

\acknowledgments

\bibliographystyle{apj} 
%\bibliography{jetbib}

%\begin{comment}

%\end{comment}

\end{document}